\documentclass[showpacs,preprintnumbers,amsmath,amssymb,superscriptaddress,aps]{revtex4}

\usepackage{amsmath}
\usepackage{amssymb}
\usepackage{amsfonts}
\usepackage{amsthm}
\usepackage{fancyhdr}
\usepackage{fancybox}
\usepackage{graphicx}
\usepackage{natbib}
\usepackage{float}

\renewcommand{\Re}{{\rm Re}}
\renewcommand{\Im}{{\rm Im}}
\makeatletter

\newcommand{\ket}[1]{|{#1}\rangle}

\newcommand{\bra}[1]{\langle{#1}|}
\newcommand{\bkt}[2]{\langle{#1}|{#2}\rangle}

\newcommand{\wv}[2]{\langle{#1}\rangle_{#2}}

\begin{document}
\title{Bayesian Interpretation of Weak Values}
\author{Akio Hosoya}\email{ahosoya.bongo10@gmail.com}
 \affiliation{Department of Physics, Tokyo Institute of
 Technology, Tokyo, Japan}
\date{\today}

\begin{abstract}
The real part of the weak value is identified as the conditional Bayes probability through the quantum analog of the Bayes relation.
We present   an explicit protocol to get  the the weak values  in a simple Mach-Zehnder interferometer model and derive the formulae
for the weak values in terms of the experimental data consisting of the positions and momenta of detected photons on the basis of the quantum Bayes relation.
The formula gives a way of tomography of the initial state almost without disturbing it in the weak coupling limit.

 \end{abstract}

\maketitle
\section{Introduction}
 The concept of weak value was originally proposed by Aharonov and his collaborators~\cite{AAV,AR} in terms 
of the weak measurement. Many experiments have been carried out~\cite{RSH,PCS,RLS,Pryde,HK,RESCH}. 
On the basis of the weak measurement, a formal theory of the weak value has also been developed~\cite{MJP}.
However, we believe that the weak values have to be studied on its own right  ~\cite{HK2}, although its measurability is utmost 
important.
 
The weak value of an observable $\hat{A}$  is defined by
\begin{equation}
_{z}\wv{\hat{A}}{\psi}^{w}:=\frac{\bra{z}\hat{A}\ket{\psi}}{\bkt{z}{\psi}}, \ \bkt{z}{\psi} \neq 0, 
\label{defi}
\end{equation}
where $\ket{\psi}$ and $\bra{z}$ are the initial and final states.
In this work we specifically consider the case that $\hat{A}=\ket{a}\bra{a}$ is a projection operator. Note that in this case the weak value becomes a portion of the
partial amplitudes~\cite{FEYNMAN}
\begin{equation}
_{z}\wv{\hat{A}}{\psi}^{w}=\frac{\bkt{z}{a}\bkt{a}{\psi}}{\sum_{m}\bkt{z}{m}\bkt{m}{\psi}}.
\label{portion}
\end{equation}

 It is thus tempting to to interpret the weak value of the projection operator $\ket{a}\bra{a}$ as a conditional probability for a particle initially in the state $\ket{\psi}$ goes through
 a particular intermediate state $\ket{a}$ out of the  possible intermediate states $\ket{m}=\ket{a},\ket{b},\dots$ and is detected in the final state $\ket{z}$.
There have been debates over the possibility of the  probabilistic interpretation of the weak value, which some people do not like because it can be negative or even complex.
   It is probably fair to say that the  probabilistic interpretation is far from compelling, though consistent with the Kolomogorov axioms except the positivity of the probability.
   In this work we add a farther argument that supports the  probabilistic interpretation of the weak values on the basis of Bayesian statistics.
   
    In Sec II, we give a very brief introduction of the Bayes relation and the present status of quantum Bayesian statistics. We derive a quantum version of Bayes relation identifying the
   weak values as the Bayes conditional probabilities. Using a simple Mach-Zehnder interferometer model, we present in Sec III an example of the weak measurements to evaluate the weak values.
   Sec IV is devoted to summary and discussion.
 %%%%%%%%%%%%%%%%%%%%%%%%%%%%%%  
   \section{Classical and quantum Bayesian probabilities}
 %%%%%%%%%%%%%%%%%%%%%%%%%%%%%%
   The Bayesian probability is a degree of likeliness in contrast to the standard frequency based probability theory. The key equation in the Bayesian probability theory~\cite{PRESS} is the Bayes relation
   for the conditional probabilities:
 \begin{equation}
 P(A|Z)=\frac{P(Z|A)P(A)}{P(Z)},
 \label{Bayes}
\end{equation}
where $P(A)$ and $P(Z)$ are probabilities for events $A$ and $Z$ to occur, respectively. The $P(A|Z)$ ($P(Z|A)$) is the conditional probability for $A$ ($Z$) to occur under the condition $Z$ ($A$).
Normally it is assumed that $P(A)$ is prior given, while the conditional probability $P(Z|A)$  is obtained by measurements and $P(Z)$ is calculated to be $P(Z)=\sum_{A}P(Z|A)P(A)$.
Then we can obtain by (\ref{Bayes}) the probability $P(A|Z)$.
One of the widely used practical application is to update  the probability of $A$ by replacing the a priori probability $P(A)$ by $P(A|Z)$ with the new experience $Z$.

 Alternatively, we may interpret  (\ref{Bayes}) as follows. The $P(A)$ is the a priori probability for a hypothesis $A$ as before and the $P(Z|A)$ is suposingly obtained by observing $Z$ under the hypothesis $A$.
The left hand side $ P(A|Z)$ is the probability to retrospectively guess the hypothesis $A$ by the data $Z$. Note that this implicitly contains a causal interpretation that $A$ causes $Z$.
%%%%%%%%%%%%%%%%
\begin{figure}[t]
\begin{center}
\includegraphics[width=40mm]{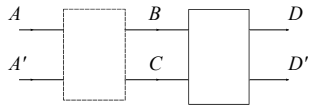}
%\caption{\label{Fig1}}
\caption[Fig.1]
{A simple Bayesian network.  $(A,A')$ and $(D,D')$ are the input and output states of a particle,while $(B,C)$ are the two possible intermediate states.
}
\end{center}
\end{figure}
%%%%%%%%%%%%%%%%

  To add a flesh to the second interpretation, consider a simple Bayesian network shown in Fig.1. A particle comes into either $B$ or $C$ with the probability $P(B)$ and $P(C)$, respectively, and then is detected at either $D$ or $D'$.
  The Bayes relation (\ref{Bayes}) reads,
   \begin{equation}
 P(B|D)=\frac{P(D|B)P(B)}{P(D|B)P(B)+P(D|C)P(C)},
 \label{Bayesianet}
\end{equation}
 and similar expressions for $P(C|D)$, where $P(D|B)$ and $P(D|C)$ are the conditional probabilities for the particle will be detected when it comes into $B$ and $C$, respectively.

The Bayesian update method in the first interpretation has already been directly applied to the state estimation in the context of quantum measurement theories by several authors. Jones~\cite{J} proposed the quantum Bayesian
 update of a pure state and later Schack, Brun and Caves extended it to a certain class of mixed states ~\cite{SBC}, in which $A$ in (\ref{Bayes})  is the initial density operator  and $Z$ is the measurement outcome by a POVM
 in the generalized measurement.
 The formula for the updated density operator looks like the Bayesian relation, which they call the quantum Bayesian relation.
 
  In the present work, however, we fix the initial state once and for all,  and concentrate on the likeliness of the intermediate state from the probability to obtain the final state. 
  Here lurks a subtle point in quantum mechanics: the conditioning the intermediate state keeping the initial state intact. 
  We assume that the likeliness of the intermediate state is given by the weak value of the projection
 operator to the intermediate state for given initial and final states. In a sense we propose a quantum analog of classical hypothesis guessing rather than the tomography of the initial state. In the quantum case the causality has to be carefully treated; the seemingly counter-factual conditioning of the intermediate state is realized
 by a tiny interaction of the measured system and the meter, a key ingredient of the weak measurement.
 
  From our point of view, this use of the Bayesian relation is  more fundamental than the tomography in quantum mechanics. We have to caution the reader that we stick to the standard quantum mechanics not trying to reconstruct quantum mechanics on the basis of Bayesian probability theory~\cite{Fuchs}.
 
%%%%%%%%%%%%%%%%%%%%%%%%%%%%%%%
  \section{Quantum Bayesian relation}
%%%%%%%%%%%%%%%%%%%%%%%%%%%%%%%
Let the initial state be $\ket{\psi}$ and the final (post-selected) state be $\ket{z}$ and consider the weak value\cite{AAV} of the projection operator $\hat{A}=\ket{a}\bra{a}$
defined by
\begin{equation}
\frac{\bra{z}\hat{A}\ket{\psi}}{\bkt{z}{\psi}}, (\bkt{z}{\psi}\neq 0)
 \label{weakvalue}
\end{equation}
By a simple manipulation, we obtain
\begin{eqnarray}
\frac{\bra{z}\hat{A}\ket{\psi}}{\bkt{z}{\psi}}&=\frac{\bra{z}(\ket{a}\bra{a})\ket{\psi}}{\bkt{z}{\psi}}\\
=\frac{\bkt{\psi}{z}\cdot \bkt{z}{a}\bkt{a}{\psi}}{|\bkt{z}{\psi}|^2}&=\frac{\bkt{\psi}{z}\bkt{z}{a}}{\bkt{\psi}{a}}\cdot\frac{|\bkt{a}{\psi}|^2}{|\bkt{z}{\psi}|^2}\\
=\frac{\bra{\psi}\hat{Z}\ket{a}}{\bkt{\psi}{a}}\cdot\frac{P(a)}{P(z)}&=[\frac{\bra{a}\hat{Z}\ket{\psi}}{\bkt{a}{\psi}}]^{*}\cdot\frac{P(a)}{P(z)}
 \label{weakvalue1}
\end{eqnarray}
where $\hat{Z}$ is the projection operator $\hat{Z}=\ket{z}\bra{z}$ and $P(a)=|\bkt{a}{\psi}|^2$ and $P(z)=|\bkt{z}{\psi}|^2$ are probabilities to get $a$ and $z$, respectively according to the
Born rule. We see that the two kinds of the weak values $\frac{\bra{z}\hat{A}\ket{\psi}}{\bkt{z}{\psi}}$ and $\frac{\bra{a}\hat{Z}\ket{\psi}}{\bkt{a}{\psi}}$ are
related by the Bayese-like relation. Note that the factor $\frac{\bra{\psi}\hat{Z}\ket{a}}{\bkt{\psi}{a}}$ on the right hand is the complex conjugate of the weak value
$\frac{\bra{a}\hat{Z}\ket{\psi}}{\bkt{a}{\psi}}$. This is suggestive in the probabilistic interpretation because the complex conjugation implies the time reversal in quantum mechanics.

 It is thus tempting to identify the real part of the weak value $\frac{\bra{z}\hat{A}\ket{\psi}}{\bkt{z}{\psi}}$ as the conditional probability $P(z|a)$ to arrive at
the state $\ket{z}$ via the state $\ket{a}$ and similar to the weak value, $\frac{\bra{a}\hat{Z}\ket{\psi}}{\bkt{a}{\psi}}$. 

Writing $P(z|a)=Re[\frac{\bra{z}\hat{A}\ket{\psi}}{\bkt{z}{\psi}}]$ and
$P(a|z)=Re[\frac{\bra{a}\hat{Z}\ket{\psi}}{\bkt{a}{\psi}}]$, we end up with the Bayes relation:
 \begin{equation}
 P(a|z)=\frac{P(z|a)P(a)}{P(z)}.
 \label{QBayes}
\end{equation}
We use
\begin{equation}
P(z)= \sum_{a}P(z|a)P(a)
 \label{sum}
\end{equation}
in (\ref{QBayes}) and assume $P(z|a)$ is experimentally accessible.

The choice of the real part rather than e.g., the absolute value seems unique assuming the linearity of the conditional probability: $P(a|x)+P(a|y)=P(a|x+y)$. 
If one prefers an axiomatic description, one can prove this replacing the fifth axiom in the previous paper\cite{HK2} by the Bayes relation (\ref{QBayes}
). 

We remark that the weak value can also be expressed as 
\begin{eqnarray}
\frac{\bra{z}\hat{A}\ket{\psi}}{\bkt{z}{\psi}}=\frac{\bra{\psi}\hat{Z}\hat{A}\ket{\psi}}{P(z)},
 \label{joint}
\end{eqnarray}
which suggests that $\Re\bra{\psi}\hat{Z}\hat{A}\ket{\psi}$ is the joint probability for $\ket{a}$ and then $\ket{z}$ to occur provided that the final state is $\ket{z}$.
However, caution has to be paid, because the projection operators $\hat{Z}$ and $\hat{A}$ are not in general commutable, i.e., $\hat{Z}\hat{A}\neq \hat{A}\hat{Z}$. This non-commutativity makes the weak value
complex in general. 

The imaginary part of the weak value is thus given by
\begin{eqnarray}
\Im[\frac{\bra{z}\hat{A}\ket{\psi}}{\bkt{z}{\psi}}]=\frac{1}{2i}\frac{\bra{\psi}[\hat{Z},\hat{A}]\ket{\psi}}{P(z)}.
 \label{imaginary}
\end{eqnarray}
Therefore the imaginary part of the weak value is the lower bound of the uncertainty relation
\begin{eqnarray}
|\Im[\frac{\bra{z}\hat{A}\ket{\psi}}{\bkt{z}{\psi}}]|\leq\frac{\Delta Z\cdot\Delta A}{P(z)},
 \label{uncertainty}
\end{eqnarray}
where $\Delta Z$ and $\Delta A$ are quantum fluctuation of $Z$ and $A$ in the state $\ket{\psi}$.

Note also that $\arg(\frac{\bra{z}\hat{A}\ket{\psi}}{\bkt{z}{\psi}})=\arg(\bra{\psi}\hat{Z}\hat{A}\ket{\psi})$ is the geometric phase for the loop of the state change: $\ket{\psi}\rightarrow \ket{a}\rightarrow \ket{z}\rightarrow \ket{\psi}$
~\cite{BZ}.

 So far is a formal and mathematical discussion. Suppose we interpret the weak values $\frac{\bra{z}\hat{A}\ket{\psi}}{\bkt{z}{\psi}}$ and $\frac{\bra{a}\hat{Z}\ket{\psi}}{\bkt{a}{\psi}}$  as conditional probabilities.
 What probability does it  precisely mean?  We propose that the former weak value is the probability for the state to pass through $\ket{a}$ provided that the initial state is $\ket{\psi}$ and the final state is $\ket{z}$ as some of the
 advocates of the probabilistic interpretation of weak values claim. The latter is a new face, because the state $\ket{a}$ is the final state rather than the intermediate state. Our  interpretation of it is the counter factual probability distribution that we would obtain $z$ if the state passed
 through the state $\ket{a}$. To demonstrate that this interpretation works, we present a simple Mach-Zehnder model in the following section.
 %%%%%%%%%%%%%%%%% 
%%%%%%%%%%%%%%%%% 
\section{Weak measurement in a simple  Mach-Zehnder model}
%%%%%%%%%%%%%%%%
 Consider the standard Mach-Zehnder interferometer depicted in Fig.1 and assume that the incident photon enters the interferometer from the left-top as a superposition 
$
\ket{\psi}=\beta\ket{B}+\gamma \ket{C},
$
where $\beta$ and $\gamma$ are given but not necessarily known complex coefficients. The beam splitters make the following unitary transitions,
$
\ket{A}\rightarrow \frac{1}{\sqrt{2}}[\ket{B}+\ket{C}],\;\;\ket{A'}\rightarrow \frac{1}{\sqrt{2}}[\ket{B}-\ket{C}],\;\;
\ket{B}\rightarrow \frac{1}{\sqrt{2}}[\ket{D}-\ket{D'}],\;\;\ket{C}\rightarrow \frac{1}{\sqrt{2}}[\ket{D}+\ket{D'}].
$
and therefore the sequential transition becomes
$
\ket{A}\rightarrow \ket{D},\;\;\ket{A'}\rightarrow -\ket{D'}.
$
%%%%%%%%%%%%%%%%
\begin{figure}[t]
\begin{center}
\includegraphics[width=40mm]{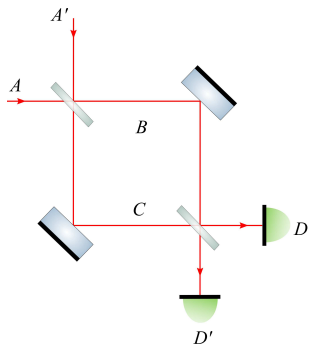}
%\caption{\label{Fig1}}
\caption[Fig.2]
{A standard Mach-Zehnder interferometer consisting of the two beam splitters and the two mirrors. The path-states $(A,A'),(B,C)),(D,D')$ correspond to the states in Fig.1.
A photon is injected as a superposition of $A$ and $A'$ and detected at $D$ or $D'$.
}
\end{center}
\end{figure}
%%%%%%%%%%%%%%%%

 We interpret the weak value $\Re(\frac{\bra{D}\hat{B}\ket{\psi}}{\bkt{D}{\psi}})$ as the a posteriori probability for the photon to have taken the path $B$ looking backward in time provided that the photon is found in the port $D$, which
 can be extracted by the weak measurement in the sense that the probe wave function in the coordinate representation is displaced by the amount proportional to the real part of the weak value. The imaginary part shows up
as the displacement in the momentum representation.~\cite{AAV}.
 
   More precisely, insert a slightly tilted thin slide glass in the path $B$  in the Mach-Zehnder interferometer (Fig.3). This introduces the interaction between the original optical system and the newly introduced
   degree of freedom $z$ perpendicular to the optical axis, which we call the probe\cite{AR}. 
   %%%%%%%%%%%%%%%%%%%%%%%%%%%%%%%%
   %%%%%%%%%%%%%%%%%%%%%%%%%%%%%%%
 \begin{figure}[t]
 \begin{center}
\vskip 3 cm
\includegraphics[width=40mm]{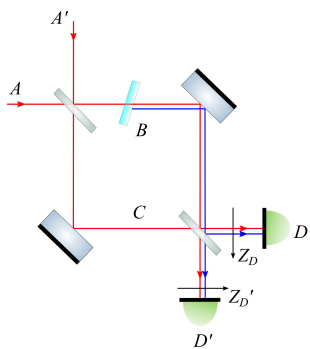}
   \caption[Fig.3]
{A Mach-Zehnder interferometer with a tilted thin slide glass inserted in the path $B$. A photon detected by $D$ ($D'$) is  deflected by $Z_D$ ($Z_D'$) in the direction perpendicular to the optical axis.
}
\label{MZ2}
\end{center}
\end{figure}
%%%%%%%%%%%%%%%%%%%%%%%%%%%%%%%%
The thin slide glass refracts the photon off the optical axis by $g$ only if it passes through $B$. The interaction Hamiltonian
   is given by
   \begin{eqnarray}
H=g\hat{B}p_z\delta(t),
 \label{Hamiltonian}
\end{eqnarray}
where $p_z$ is the momentum conjugate to the perpendicular displacement $z$. According to the standard weak measurement protocol, we post select the $D$-port
to count the number of photons detected at $z$. Let the initial probe wave function be $f(z)$. The probability distribution is then predicted for a small $g$ as
\begin{eqnarray}
Pr(z,D)=|\bkt{\psi}{D}|^2|f(z-g\cdot\Re(\frac{\bra{D}\hat{B}\ket{\psi}}{\bkt{D}{\psi}}))|^2.
\label{distribution}
\end{eqnarray}
% By carrying out the photon detection at different positions $z$ many times we can obtain the distribution and then the average shift   $gRe(\frac{\bra{D}\hat{B}\ket{\psi}}{\bkt{D}{\psi}})$ of the probe, from which we can extract
%the weak value $Re(\frac{\bra{D}\hat{B}\ket{\psi}}{\bkt{D}{\psi}})$ dividing the shift by the known parameter $g$. So far is by now the standard thing\cite{AR}. Note that the standard argument discarded the information of the
%prefactor $|\bkt{\psi}{D}|^2$, which will be taken care below in the evaluation of the other weak value $Re(\frac{\bra{\psi}\hat{D}\ket{B}}{\bkt{\psi}{D}})$.
 
% Let us now consider the quantity
% \begin{eqnarray}
%z_D=\int dz\;zPr(z,D),
%\label{xD}
%\end{eqnarray} 
%which is experimentally accessible by measuring the vertical distance of each detected particle and the sum them up and divide the result by the number of incident particles. Theoretically the result should be
%\begin{eqnarray}
%z_D=g|\bkt{\psi}{D}|^{2}Re(\frac{\bra{D}\hat{B}\ket{\psi}}{\bkt{D}{\psi}}).
%\label{xDth}
%\end{eqnarray} 

%%%%%%%%%%%%%%%%% 
%\section{Wave-particle duality}
%%%%%%%%%%%%%%%%
  Following the standard weak measurement formulation the real part of weak value $\frac{\bra{D}\hat{B}\ket{\psi}}{\bkt{D}{\psi}}$ can be obtained by the shift of the average of the probe position:
 \begin{eqnarray}
g\cdot \Re[\frac{\bra{D}\hat{B}\ket{\psi}}{\bkt{D}{\psi}}]=\frac{Z_{D}}{N_D},
\label{shift}
\end{eqnarray} 
if we know the parameter $g$.
Here $Z_{D}=z_{1}+\cdots z_{N_D}$ is the sum of the z-coordinates of the detected  $N_D$ photons at the D-port. The right hand side can also be obtained as the center of the distribution by classical light rather than by the photon counting much more easily.

 In what follows, using the same Mach-Zehnder interferometer model, we show a protocol to extract the real part of the other weak value $\frac{\bra{\psi}\hat{D}\ket{B}}{\bkt{\psi}{B}}$ on the right hand side of the Bayes relation:
  \begin{eqnarray}
\Re(\frac{\bra{\psi}\hat{D}\ket{B}}{\bkt{\psi}{B}})=\Re(\frac{\bra{D}\hat{B}\ket{\psi}}{\bkt{D}{\psi}})\cdot\frac{|\bkt{D}{\psi}|^2}{|\bkt{B}{\psi}|^2}.
 \label{MZ Bayes}
\end{eqnarray}
 in terms of experimentally accessible quantities. 
  Note that $|\bkt{D}{\psi}|^{2}=\frac{N_D}{N}$, where $N$ is the number of incident photons and $N_D$ is the number of detected photons at $D$ port and that $|\bkt{B}{\psi}|^{2}=\bra{\psi}\hat{B}\ket{\psi}=\sum_{d=D,D'}\bkt{\psi}{d}\bra{d}\hat{B}\ket{\psi}=\sum_{d=D,D'}|\bkt{\psi}{d}|^{2}\frac{\bra{d}\hat{B}\ket{\psi}}{\bkt{d}{\psi}}$. Therefore, we obtain
 
 \begin{eqnarray}
\Re(\frac{\bra{\psi}\hat{D}\ket{B}}{\bkt{\psi}{B}})=\frac{Z_{D}}{Z_{D}+Z_{D'}},
\label{ratio1}
\end{eqnarray} 
where $Z_{D}=z_{1}+\cdots z_{N_D}$ and $Z_{D'}=z'_{1}+\cdots z'_{N_{D'}}$ are the sums of the z-coordinates of the detected  $N_D$ photons at the $D$-port and that of the $D'$-port, respectively.
We note that all the informations of the coordinates of the detected photons are involved in (\ref{ratio1}),while the expression (\ref{shift}) uses only the partial data obtained at $D$-port,while the data at the $D'$-port are thrown away.

The original weak value $\Re[\frac{\bra{D}\hat{B}\ket{\psi}}{\bkt{D}{\psi}}]$ can be re-expressed via Bayes' rule by
 \begin{eqnarray}
\Re[\frac{\bra{D}\hat{B}\ket{\psi}}{\bkt{D}{\psi}}]=\Re[\frac{\bra{\psi}\hat{D}\ket{B}}{\bkt{\psi}{B}}]\cdot\frac{|\bkt{B}{\psi}|^2}{|\bkt{D}{\psi}|^2}\\
=\frac{\frac{Z^{B}_{D}}{Z^{B}_{D}+Z^{B}_{D'}}P_{B}}{\frac{Z^{B}_{D}}{Z^{B}_{D}+Z^{B}_{D'}}P_{B}+\frac{Z^{C}_{D}}{Z^{C}_{D}+Z^{C}_{D'}}P_{C}}=\frac{Z^{B}_{D}}{Z^{B}_{D}+Z^{C}_{D}}.
 \label{MZ Bayes3}
\end{eqnarray}
Here the superfix $B$ ($C$) to the coordinate $Z$ indicates the z-coordinate shift of the photon when the slide glass is inserted in the path $B$ ($C$). Note that the two expressions for the weak values do not contain
the coupling parameter $g$  in the weak measurement but only the experimentally obtainable quantities $Z^{B}_{D}, Z^{B}_{D'}, Z^{C}_{D}$ and $Z^{C}_{D'}$.

 To get some insight it would be helpful to see a numerical example: $\beta=\sqrt{1/5},\; \gamma=-\sqrt{4/5}$, which gives a negative weak value $\Re[\frac{\bra{D}\hat{B}\ket{\psi}}{\bkt{D}{\psi}}]=\frac{\beta}{\beta+\gamma}=-1$ and the
prior probabilities $P_{B}=\frac{1}{2}(\beta+\gamma)^2=1/10$ and $P_{C}=\frac{1}{2}(\beta-\gamma)^2=9/10$. These numerical values can be realized for example by the data $Z^{B}_{D}=-1,\;Z^{C}_{D}=2$ in some
unit of length.
The negative weak value can intuitively understood by the negative deflection $Z^{B}_{D}=-1$ relative to the total deflection $Z^{B}_{D}+Z^{C}_{D}=1$.

          The imaginary part of the weak value ~\cite{MJP} is obtained by the momentum shift of the probe as
\begin{eqnarray}
2g\cdot Var(p) \Im[\frac{\bra{D}\hat{B}\ket{\psi}}{\bkt{D}{\psi}}]=\frac{P^{B}_{D}}{N_D},
\label{momentumshift}
\end{eqnarray} 
where $P^{B}_{D}=p_{1}+p_{2}+\dots +p_{N_{D}}$ is the sum of the momentum in the z-direction of the probe and the $Var(p)$ is the variance of the momentum distribution of the probe. Combining this with (\ref{uncertainty}) we see an inequality:
          \begin{eqnarray}
\frac{P^{B}_{D}}{2g\cdot Var(p) N}\leq \Delta B\cdot\Delta D.
\label{momentumshift}
\end{eqnarray}

 Alternatively, we may combine the real and imaginary parts of the weak value as
 \begin{eqnarray}
g\cdot \frac{\bra{D}\hat{B}\ket{\psi}}{\bkt{D}{\psi}}=\frac{\xi^{B}_{D}}{N_D},
\label{compexshift}
\end{eqnarray} 
where 
 \begin{eqnarray}
\xi^{B}_{D}:=Z^{B}_{D}+i\frac{P^{B}_{D}}{2Var(p)}
\label{compexshift2}
\end{eqnarray} 
and similar expressions replacing $B$ by $C$ and or $D$ by $D'$. Here $Z^{B}_{D}=z_{1}+\cdots z_{N_D}$ and $P^{B}_{D}=p_{1}+p_{2}+\dots +p_{N_{D}}$.

The complex Bayes relation reads
  \begin{eqnarray}
(\frac{\bra{\psi}\hat{D}\ket{B}}{\bkt{\psi}{B}})^{*}=\frac{\bra{D}\hat{B}\ket{\psi}}{\bkt{D}{\psi}}\cdot\frac{P(D)}{P(B)},
 \label{complexBayes}
\end{eqnarray}
where $P(D)=\frac{N_{D}}{N}$ as before and
  \begin{eqnarray}
P(B)=\sum_{d=D,D'}\frac{\bra{d}\hat{B}\ket{\psi}}{\bkt{d}{\psi}}P(d)\\
=\frac{1}{g}(\frac{\xi^{B}_D}{N_D}\frac{N_D}{N}+
\frac{\xi^{B}_{D'}}{N_{D'}}\frac{N_{D'}}{N})\\
=\frac{1}{gN}(\xi^{B}_{D}+\xi^{B}_{D'}).
 \label{PB}
\end{eqnarray}
Therefore, we obtain an expression for the weak value
  \begin{eqnarray}
(\frac{\bra{\psi}\hat{D}\ket{B}}{\bkt{\psi}{B}})^{*}=\frac{\xi^{B}_{D}}{\xi^{B}_{D}+\xi^{B}_{D'}}=:\eta^{B},
 \label{eta}
\end{eqnarray}
 in terms of the complex shifts $\xi^{B}_D$ and $\xi^{B}_{D'}$ by the insertion of the slide glass in the path $B$.
 $\xi's$ are experimentally accessible in the measurement of the coordinate and momentum shifts from the optical axis of the detected photon when
 we inject photons one by one into the Mach-Zehnder system.
 
 Further we see that
\begin{eqnarray}
 \frac{\bra{D}\hat{B}\ket{\psi}}{\bkt{D}{\psi}}=\frac{\eta^{B}P(B)}{\eta^{B}P(B)+\eta^{C}P(C)}=\frac{\xi^{B}_D}{\xi^{B}_D+\xi^{C}_D}.
 \end{eqnarray}
holds which is equivalent to  (\ref{MZ Bayes3}) and (\ref{compexshift}).

 Comparing this with the theoretical weak value $ \frac{\bra{D}\hat{B}\ket{\psi}}{\bkt{D}{\psi}}=\frac{\beta}{\beta+\gamma}$, the portion of the partial amplitudes,we obtain a simple formula for the tomography of the initial state $\ket{\psi}=\beta\ket{B}+\gamma\ket{C}$
 as 
 \begin{eqnarray}
 \frac{\gamma}{\beta}=\frac{\xi^{C}_{D}}{\xi^{B}_{D}}
 \end{eqnarray}
by the measuring the coordinates and momenta of the detected photons in the weak measurements (See (\ref{compexshift2})).
At this stage we can confirm the claim of AAV~\cite{AAV} that the weak measurement gives the full information of the initial state almost without disturbing it, because the ratio $\frac{\xi^{C}_{D}}{\xi^{B}_{D}}$ can be defined
in the weak coupling limit $g\rightarrow 0$.

Similarly from $\frac{\bra{D'}\hat{B}\ket{\psi}}{\bkt{D'}{\psi}}=\frac{\beta}{\beta-\gamma}$,we have $\frac{\gamma}{\beta}=-\frac{\xi^{B}_{D'}}{\xi^{C}_{D'}}$.
This leads to a relation among the "complex shifts",
 \begin{eqnarray}
 \frac{\xi^{B}_{D}}{\xi^{C}_{D}}=-\frac{\xi^{B}_{D'}}{\xi^{C}_{D'}},
 \end{eqnarray}
 which may be useful to check the consistency of the experimental results.

%%%%%%%%%%%%%%%%% 
\section{Summary and discussion}
%%%%%%%%%%%%%%%% 
A quantum version of Bayes' rule 
\begin{eqnarray}
\frac{\bra{D}\hat{B}\ket{\psi}}{\bkt{D}{\psi}}=(\frac{\bra{\psi}\hat{D}\ket{B}}{\bkt{\psi}{D}})^{*}\cdot\frac{|\bkt{B}{\psi}|^2}{|\bkt{D}{\psi}|^2}
\label{qbayes}
\end{eqnarray}
has been derived.
We interpret the weak value $\Re(\frac{\bra{D}\hat{B}\ket{\psi}}{\bkt{D}{\psi}})$ in the quantum Bayes relation (\ref{QBayes}) as  the likeliness of the path $\hat{B}$ in the past when the photon is detected in a definite state $\ket{D}$ and the other weak value $Re(\frac{\bra{\psi}\hat{D}\ket{B}}{\bkt{\psi}{B}})$ as the probability distribution of the final state $\ket{D}$ for a counter-factual intermediate state $\ket{B}$.
  An example of protocols to measure the two  weak values is shown in a  Mach-Zehnder interferometer model.
  The results are 
    \begin{eqnarray}
   \nonumber
  (\frac{\bra{\psi}\hat{D}\ket{B}}{\bkt{\psi}{B}})^{*}&=\frac{\xi^{B}_{D}}{\xi^{B}_{D}+\xi^{B}_{D'}},\\ 
  \frac{\bra{D}\hat{B}\ket{\psi}}{\bkt{D}{\psi}}&=\frac{\xi^{B}_D}{\xi^{B}_D+\xi^{C}_D}.
 \label{summary}
 \end{eqnarray}
 where the complex shifts $\xi^{B}_{D},\xi^{B}_{D'},\xi^{C}_{D}$ and $\xi^{C}_{D'}$ are defined for each path ($B$ or $C$) and detected port $D$ or $D'$ in a way that $\xi^{B}_{D}:=Z^{B}_{D}+i\frac{P^{B}_{D}}{2Var(p)}$ etc..
 The equations (\ref{qbayes}) and (\ref{summary}) are the main results of the present work.
 
 The complex shifts are constrained by a relation:$ \frac{\xi^{B}_{D}}{\xi^{C}_{D}}=-\frac{\xi^{B}_{D'}}{\xi^{C}_{D'}}$.
  Although we have analyzed a specific Mach-Zehnder model, it seems that the present argument can be straightforwardly generalized to other cases.

  There have been debates whether the weak measurement is really quantum phenomenon, because the result can also be understood in classical wave mechanics.
 It is interesting to point out that the wave-particle duality can be seen in the Bayes relation in an experimentally verifiable way  by a simple Mach-Zehnder interferometer.  The left hand side of the Bayes relation is obtained by the shift of the probe distribution, which
is also possible even by analog classical light while the right hand side can be constructed by  a set of measured distances in the z-direction of each detected photon, which cannot  be collectively achieved  by classical light.

  We have seen another supporting evidence of the consistency of the probabilistic interpretation of the weak value. On the basis of the quantum Bayesian relation combined with the model of the weak measurement, we have  obtained  a simple
  formula for the weak value in terms of the positions of the detected photons. 
  %The similarity of (\ref{MZ Bayes3}) and  (\ref{Bayesianet}) is striking, if we replace $N_D,N'_D$ by the z-coordinates $Z_D,Z'_D$. It is then tempting to consider
 % an interpolating quantity:$Z^{q}_D=z^{q}_{1}+\cdots z^{q}_{N_D},\;\; 0\leq q< +\infty$.
 %%%%%%%%%%%%%%%

\end{document}